\documentclass[preprint,amsmath,amssymb,aps,prd,showpacs]{revtex4-1}
\usepackage{graphicx}
\usepackage{dcolumn}
\usepackage{bm}
\newcommand{\re}{\ensuremath{\mathrm{Re}}}
\newcommand{\arccosh}{\ensuremath{\mathrm{arccosh}}}
\begin{document}
\title{Complex saddles in the Gross-Witten-Wadia matrix model}

\author{Gabriel \'Alvarez}
\email{galvarez@ucm.es}
\author{Luis Mart\'{\i}nez Alonso}
\email{luism@ucm.es}
\affiliation{
	        Departamento de F\'{\i}sica Te\'orica II,
                Facultad de Ciencias F\'{\i}sicas,
                Universidad Complutense,
                28040 Madrid, Spain
}

\author{Elena Medina}
\email{elena.medina@uca.es}
\affiliation{
                Departamento de Matem\'aticas,
                Facultad de Ciencias,
                Universidad de C\'adiz,
                11510 Puerto Real, Spain
}

\date{\today}
\begin{abstract}
We give an exhaustive characterization of the complex saddle point configurations of the Gross-Witten-Wadia
matrix model in the large-$N$ limit. In particular, we characterize the cases in which the saddles accumulate
in one, two, or three arcs, in terms of the values of the coupling constant and of the fraction of the total unit
density that is supported in one of the arcs, and derive an explicit condition for gap closing associated to
nonvacuum saddles. By applying the idea of large-$N$ instanton we also give direct analytic derivations
of the weak-coupling and strong-coupling instanton actions.
\end{abstract}
\pacs{12.38.Aw; 02.10.Yn}
\maketitle
\section{Introduction}
The $1/N$ expansion was introduced by {}'t~Hooft for the study of $SU(N)$ color theories of
the strong interaction~\cite{TH74}. {}'t~Hooft showed that the leading contribution to each order in the
now-called  {}'t~Hooft parameter $\lambda=g^2 N$ comes from planar diagrams, where $g^2$ is the gauge
coupling and $\lambda$ is kept fixed as $N\to\infty$. Since then, the $1/N$ expansion has been used
in many different settings, and it has been realized the role played by the two parameters $N$ and $\lambda$
not just at the perturbative level, but to describe fluctuations around saddle points and more general
non-perturbative phenomena. The analysis of these non-perturbative phenomena requires methods beyond
classical asymptotics, and leads to new results in matrix models, gauge theories, supersymmetric
gauge theories and string theories~\cite{MA08, MA08b, AN15}.

Non-perturbative effects in matrix models were first studied by David~\cite{DA91,DA93} in the critical case
and using the double-scaling limit method, but it was soon realized that these effects are also worth studying
in matrix models off criticality. Among the key realizations were (i) that the set of saddle points of a matrix model
allows us to describe the nonperturbative corrections to the free energy in terms of large-$N$ instantons,
i.e., migration of eigenvalues among different pieces of the eigenvalue density support,
and (ii) the role played by complex saddles even in cases where the original partition functions were
sums over real configurations~\cite{BO00,MA08,MA08b,EY09,MAR09,HM11,DD11}.

In a recent paper Buividovich, Dunne, and Valgushev~\cite{BD16} applied these ideas to a detailed numerical
study of the complex saddle points in the Gross-Witten-Wadia (GWW) unitary matrix model, which is equivalent
to a simple 2D lattice gauge theory~\cite{GW80,WA80,WA12}. Their numerical calculations go beyond
the well-known results on the vacuum configuration (both in the weak- and strong-coupling phases) to
discover new configurations of complex saddle points which can be interpreted as non-perturbative
effects.

The purpose of the present paper is to provide an exhaustive classification of these configurations
in the large-$N$ limit, as well as to give independent analytic derivations of the corresponding weak-coupling and
strong-coupling instanton actions that do not rely on a trans-series ansatz in the string equation~\cite{MA08b}.

In 2005 Mizoguchi~\cite{MI05} used a change of variables to transform the partition function of the unitary GWW
matrix model into the partition function of an hermitian matrix model defined on the real line.

Our strategy it to use directly the matrix eigenvalues as integration variables, thus keeping
the simpler form of the potential and in effect allowing us to write and solve a system of equations
similar to those used in Refs.~\cite{AL13,AL15b} to study the arcs that support the asymptotic
density of zeros in families of non-hermitian orthogonal polynomials.

The layout of the paper is as follows. In Sec.~\ref{sec:sad} we present this change of variables, quickly
review the saddle-point method and set up the system of equations that will eventually determine the
arcs of the eigenvalue support and the corresponding eigenvalues densities. In Sec.~\ref{sec:vac}, before
embarking on the general solution of the system, we present the simple particular solutions that correspond
to the vacuum configuration both in the weak- and in the strong-coupling regimes.  Section~\ref{sec:gs}
contains the main results of the paper, including the general solution and the classification of the possible
configurations. This classification is presented graphically in two forms: the first results from the direct
application of the method; the second, that has to be derived from the first, is simpler, physically more
meaningful, and permits direct comparison with the numerical results of Ref.~\cite{BD16}. In this section
we also give an explicit condition for the gap closing in the nonvacuum configuration in terms of the charge
fraction supported on the unit circle, as well as straightforward derivations of the weak-coupling
and strong-coupling instanton actions mentioned earlier, and some results on the limiting phase
in which all the eigenvalues have moved away from the unit circle. The paper ends with a brief Summary.
\section{Eigenvalue densities in the large $N$ limit\label{sec:sad}}
The partition function $\mathcal{Z}_N$ of the GWW unitary matrix model can be expressed in terms of the matrix 
eigenvalues $e^{-i \theta_i}$ as
\begin{equation}
	\label{umm}
	\mathcal{Z}_N
	=
	\prod_{i=1}^N
	\int_{-\pi}^{\pi}  d\theta_i
	\prod_{i<j} \sin^2\left(\frac{\theta_i-\theta_j}{2}\right)
	\exp\left(\frac{2N}{\lambda}\cos \theta_i\right),
\end{equation}
where $N$ is the dimension of the matrices and $\lambda>0$ is the {}'t~Hooft parameter. In Ref.~\cite{MI05}
Mizoguchi used the change of variables
\begin{equation}
	\label{eq:mcv}
	z_i = \tan(\theta_i/2)
\end{equation}
to write Eq.~(\ref{umm}) as the partition function of an Hermitian matrix model defined on the real line with potential
\begin{equation}
	\label{peem}
 	W_\mathrm{M}(z) = \lambda\ln(1+z^2) - \frac{2(1-z^2)}{1+z^2}.
\end{equation}
(Mizoguchi denotes the  {}'t~Hooft parameter by $\mu$ instead of $\lambda$.) However, considering the complex saddle points
found in Ref.~\cite{BD16}, we use directly the matrix eigenvalues as integration  variables, i.e.,
\begin{equation}
	z_i = \exp{(-i\theta_i)},
\end{equation}
and rewrite the GWW partition function as
\begin{equation}
	\mathcal{Z}_N = (-1)^N i^{-N^2} 2^{-N(N-1)} Z_N,
\end{equation}
where $Z_N$ is given by
\begin{equation}
	\label{hmm}	
	Z_N
	=
	\prod_{i=1}^N \int_{\Gamma} 
	  d  z_i
	\prod_{i<j}(z_i-z_j)^2
	\exp\left(-\frac{N}{\lambda}W(z_i)\right),
\end{equation}	
with $\Gamma$ being the negatively-oriented unit circle in the complex  plane, and
\begin{equation}
	\label{pee}
 	W(z) = \lambda \log z-z-\frac{1}{z}.
\end{equation}
The logarithmic function in Eq.~(\ref{pee}) is defined by $\log z=\ln|z|+{i} \arg z$ with $0\leq \arg z< 2\pi$.
The partition function in Eq.~(\ref{hmm}) determines a non-Hermitian holomorphic matrix model~\cite{LA03,FE04}
which can be analyzed in the same way  as the models recently considered in Refs.~\cite{AL13,AL15b,AL15}.

It should be noticed that Eq.~(\ref{hmm}) is a particular case of the partition function proposed
in Refs.~\cite{EG10,EG10b} to describe the Seiberg-Witten theory with $N_f=2$ flavors
\begin{equation}
	\label{SW}
 	W_{{\rm SW}}(z) = \mu_3 \log z-\frac{\Lambda_2}{2}\left(z+\frac{1}{z}\right),
\end{equation}
which in turn is a certain limiting case~\cite{EG10,EG10b} of the Penner matrix model with three logarithmic terms,
 \begin{equation}
 	\label{Pe}
 	W_{{\rm P}}(z) = m_0\log z+m_1\log (z-1)+m_2\log (z-q).
\end{equation}
\subsection{The saddle point method}
The partition function of a generic holomorphic model of the form given by Eq.~(\ref{hmm}) with an integration path $\Gamma$
can be written as
\begin{equation}
	\label{mmm}
	Z_{N}
	=
         \prod_{i=1}^N \int_{\Gamma }  d  z_i \, e^{-N^2 S_N},
\end{equation}
where the discrete action $S_N$ is 
\begin{equation}
	\label{inte}
	S_N = \frac{1}{\lambda N}\sum_i W(z_i) - \frac{1}{2 N^2}\sum_{i}\sum_{ j\neq i}\log(z_i-z_j)^2.
\end{equation}
(In case the integration path $\Gamma$ is unbounded it must lead to a convergent integral.)

The saddle point method assumes that in the large $N$ limit  the partition function $Z_N$ is dominated by the contribution
from a sequence $z_i^{(N)}$ of  saddle points  of  $S_N$ , i.e., of solutions of $\partial S_N/\partial z_i =0$,
or more explicitly, the solutions of
\begin{equation}
	\label{sa}
	\frac{N}{\lambda}W'(z_i)+\sum_{j\neq i}\frac{2}{z_j-z_i}=0.
	\quad
	i=1,\ldots,N.
\end{equation}
This is precisely the system of equations studied numerically in Ref.~\cite{BD16} for the GWW model.
Note also that the sequence $z_i^{(N)}$ should be reachable by a suitable deformation of $\Gamma$ in the analyticity
domain of $e^{-N^2 S_N}$. For later reference and as an important consequence of Eq.~(\ref{sa}),
we also recall that the discrete resolvent $\omega_N(z)$ defined by
\begin{equation}
	\label{re2}
	\omega_N(z) = \frac{1}{N}\sum_{i} \frac{1}{z-z_i^{(N)}},
\end{equation} 
satisfies the Riccati equation
\begin{equation}
	\label{ri}
	\frac{1}{N}\omega'_N(z) + \omega_N(z)^2-\frac{1}{\lambda}W'(z)\omega_N(z)
	=
	-\frac{1}{\lambda N}\sum_{i} \frac{W'(z)-W'(z_i^{(N)})}{z-z_i^{(N)}}.
\end{equation}
\subsection{The large $N$ limit: vacuum and nonvacuum configurations}
As $N\rightarrow \infty$ we expect that the sequence of saddle points accumulates over some curve $\gamma$
(made in general of a finite number $s$ of  pieces, $\gamma=\gamma_1\cup\cdots\cup\gamma_s$)   
and leads to a unit-normalized positive eigenvalue density $\rho(z)$ supported on $\gamma$:
\begin{equation}
	\label{me}
	\frac{1}{N}\sum_{i=1}^N\delta(z-z_i^{(N)}) \rightarrow \rho(z) |d  z|.
\end{equation}
However, to study both vacuum and nonvacuum configurations, me must allow for specified filling fractions $q_i$
of the total unit-normalized density to be supported on each piece of $\gamma$, i.e.,
\begin{equation}
	\label{fff}
	\int_{\gamma_i}  \rho(z)| d  z|=q_i, \quad \sum_{i=1}^{s}q_i=1.
\end{equation}

Therefore, the continuum limit $S(z)$ of the discrete action $S_N$, defined by 
 \begin{equation}
 	\label{ese}
	S[\rho]
	=\frac{1}{\lambda}\int_{\gamma}  W(z) \rho(z) |d  z|
	-
	\frac{1}{2}\int_{\gamma} | d  z| \int_{\gamma} | d  z'| \log (z-z')^2 \rho(z) \rho(z'),
\end{equation}
is assumed to satisfy the variational equation
\begin{equation}
	\label{var}
	\frac{\delta}{\delta\rho(z)}
	\left[S[\rho]+\sum_{i=1}^s\,L_i \left(q_i-\int_{\gamma_i}\rho(z)| d  z| \right)\right] = 0,
\end{equation}
where the $L_i$ are Lagrange multipliers whose physical meaning in an electrostatic interpretation
will be discussed later. Furthermore,  the definition of the function $\log(z-z')^2$ in
Eq.~(\ref{ese}) should be understood as a sum of one-sided values
\begin{equation}
    \label{logd}
    \log(z-z')^2 = \log(z_+-z')+\log(z_--z'),\quad z,z'\in \gamma,
\end{equation}
for an appropriate branch of the logarithm $\log(z-z')$.

The variational Eq.~(\ref{var}) says that the holomorphic  potential
$W_\mathrm{h}(z)$ defined by
\begin{equation}
	\label{hpo}
	W_\mathrm{h}(z)
	=
	\frac{\delta S[\rho]}{\delta\rho(z)}\\
	=
	\frac{1}{\lambda} W(z)-\int_{\gamma} \log(z-z')^2 \rho(z') |d  z'|,
\end{equation}
must be constant on each piece of the support, 
\begin{equation}
	\label{hc}
	W_\mathrm{h}(z) = L_i,\quad z\in \gamma_i,
\end{equation}
although the values $L_i$ on the different pieces $\gamma_i$ may be different.
The derivative $W'_\mathrm{h}(z)$ of the holomorphic potential is usually denoted by $y(z)$, and reads
\begin{equation}
	\label{lai0}
	y(z) = \frac{1}{\lambda} W'(z) - 2 \omega(z),
\end{equation}
where $\omega(z)$ is the continuum limit of the discrete resolvent $\omega_N(z)$,
\begin{equation}
	\label{ef}
	\omega(z) = \int_{\gamma}\frac{\rho(z') | d  z'|}{z-z'}.
\end{equation}
 (The integral in Eq.~(\ref{ef}) must be understood as a principal value.)
Note that $\omega(z)$ is the Cauchy transform of the eigenvalue density and therefore $\omega(z)$
is an analytic function of $z$ in the complex plane minus the support  $\gamma$ of the eigenvalue density, with limiting values
given by the Shokotski-Plemelj formulas
\begin{equation}
	\label{sok}
	\omega(z_+)+\omega(z_-)=2\omega(z),
	\quad
	\omega(z_+)-\omega(z_-)= -2\pi i \frac{\rho(z)}{ d  z/| d  z|},
	\quad z\in \gamma.
\end{equation}
In particular it follows that
\begin{equation}\label{yy}
y(z)=\frac{1}{2}(y(z_+)+y(z_-)),\quad z\in \gamma.
\end{equation}
\subsection{The electrostatic interpretation}
In the Coulomb gas interpretation, the saddle point condensates that determine the large-$N$ asymptotic eigenvalue
densities as given in Eq.~(\ref{me}) are considered as continuum systems of electric charges in the plane with unit charge
density $\rho(z)$ under the external potential $\re W(z)/\lambda$.

The total electrostatic energy of such configurations is given by $\mathcal{E}[\rho] = \re S[\rho]$ or, explicitly,
\begin{equation}
	\label{total}
	\mathcal{E}[\rho]
	=
	\frac{1}{\lambda}\int_{\gamma}  \re W(z) \rho(z) |d  z|
	-
	\int_{\gamma} |d  z| \int_{\gamma} |d z'| \ln|z-z'| \rho(z) \rho(z'),
\end{equation}	
and the total electrostatic potential is $U(z) = \delta \mathcal{E}[\rho]/\delta\rho(z) = \re W_\mathrm{h}(z)$ or,
explicitly,
\begin{equation}
	\label{hpot}
	U(z)
	=
	\frac{1}{\lambda} \re W(z) - \int_{\gamma} \log|z-z'|^2 \rho(z') |d  z'|.
\end{equation}
Moreover, Eqs.~(\ref{hpot}) and~(\ref{lai0}) permit to write the electric field as
\begin{equation}
	\label{elec}
	E(z)
	=
	-\frac{\partial U}{\partial x}-i\, \frac{\partial U}{\partial y}=-\overline{y(z)}.
\end{equation}
Hence, the points of electrostatic equilibrium  are those in which $y(z)$ vanishes.
In particular, taking into account Eq.~(\ref{sok}), we see that the variational Eq.~(\ref{hc}) implies
\begin{equation}
	\label{sym}
	y(z_+)+y(z_-)=0,
	\quad
	z\in \gamma_i.
\end{equation}
Hence from Eq.~(\ref{yy}) we have that $y(z)$ vanishes in $\gamma$ and therefore the points of $\gamma$
are points of electrostatic equilibrium. In fact, the real part of Eq.~(\ref{hc})  shows that the total electrostatic
potential is constant on each piece of $\gamma$,
\begin{equation}
	\label{hc2}
	U(z) = \re L_i,\quad z\in \gamma_i,
\end{equation}
In general the constants $\re L_i$ are different and  the charge density represents a critical electrostatic
charge distribution in the sense of Mart\'{\i}nez-Finkelshtein and Rakhmanov \cite{MA11}. 

In terms of the electric field, Eq.~(\ref{sym}) implies
\begin{equation}
	\label{sp}
	E(z_+)+E(z_-)=0, \quad z\in \gamma.
\end{equation}
That is to say, the electric fields at either side of $\gamma$ are opposite or, equivalently,  the forces acting on each
element of charge at $z$ from either side of $\gamma$ are equal.  This equilibrium property is the so-called
$S$-property of Stahl~\cite{ST85a,ST85b,ST86a,ST86b} and of Gonchar and Rakhmanov~\cite{GO84,GO89}.
\subsection{Equations to determine the eigenvalue density in the GWW model}
Equations~(\ref{lai0}), (\ref{sok}) and~(\ref{sym}) show the relation between the eigenvalue density $\rho(z)$ and the
limit values of $y(z)$,
\begin{equation}
	\label{mes}
	\rho(z) |  d  z| = \pm \frac{y(z_{\pm})}{2 \pi i }  d  z,\quad z\in\gamma.
\end{equation}
Thus, to calculate the eigenvalue density we attempt to characterize $y(z)$. Recall that the $\omega(z)$
in Eq.~(\ref{lai0}) for $y(z)$ is the continuum limit of  the discrete resolvent function $\omega_N(z)$,
and therefore satisfies the continuum limit of the Riccati Eq.~(\ref{ri}), the so-called Schwinger-Dyson equation,
\begin{equation}
 	\label{rl}
	\omega(z)^2-\frac{1}{\lambda}W'(z)\omega(z)
	=
	-\frac{1}{\lambda}\int_{\gamma}\frac{W'(z)-W'(z')}{z-z'}\rho(z') | d  z'|,
\end{equation}
which in terms of $y(z)$ takes the form
\begin{equation}
	\label{r1}
	y(z)^2 = \left(\frac{1}{\lambda}W'(z)\right)^2-\frac{4}{\lambda} \int_{\gamma}\frac{W'(z)-W'(z')}{z-z'}\rho(z') | d  z'|.
\end{equation}

By substituting the derivatives of the potential given in Eq.~(\ref{pee}) for the GWW in this last equation, we find that
\begin{equation}
	\label{r2}
	y(z)^2 = \left(\frac{1}{\lambda}W'(z)\right)^2 + \frac{c}{z^2} + \frac{d}{z},
\end{equation}
where
\begin{equation}
	\label{cs1}
	c = \frac{4}{\lambda}\int_{\gamma}\frac{\rho(z) | d  z|}{z},
\end{equation}
and
\begin{equation}
	\label{cs2}
	d = \lambda c +\frac{4}{\lambda}\int_{\gamma}\frac{\rho(z)| d  z|}{z^2}.
\end{equation}
Moreover, since Eq.~(\ref{lai0}) implies that
\begin{equation}
	\label{rr2}
	y(z) = -\frac{1}{\lambda}-\frac{1}{z} + O\left(\frac{1}{z^2}\right),\quad z\rightarrow \infty,
\end{equation}
Eqs.~(\ref{r2}) and~(\ref{rr2}) show that 
\begin{equation}
\label{de}
d=\frac{4}{\lambda},
\end{equation}
 and that
\begin{equation}
	\label{e1}
	y(z)^2 = \frac{1}{\lambda^2 z^4} \left(z^4+2 \lambda z^3 +((c+1)\lambda^2-2)z^2 +2 \lambda z+1\right).
\end{equation}

Consequently, the generic structure of $y(z)^2$ is
\begin{equation}
	\label{gen}
	y(z)^2 = \frac{(z-a_1)(z-a_2)(z-b_1)(z-b_2)}{\lambda^2 z^4}.
\end{equation}
Identifying coefficients in Eqs.~(\ref{e1}) and~(\ref{gen}) for $y(z)^2$ we obtain the following four equations:
\begin{eqnarray}
	a_1+a_2+b_1+b_2 & = &-2\lambda,\label{eq:sys1}\\
	a_1a_2+a_1b_1+a_1b_2+a_2 b_1+a_2 b_2+ b_1b_2 & = & (c+1)\lambda^2-2,\label{eq:sys2}\\
	a_1a_2b_1+a_1a_2b_2+a_1b_1b_2+a_2b_1b_2 & = &-2\lambda,\label{eq:sys3}\\ 
	a_1a_2b_1b_2 & = & 1.\label{eq:sys4}
\end{eqnarray}

In analogous systems for Hermitian matrix models on the real line, the $a_i$ and $b_i$ are usually referred
to as ``endpoints,'' and the eigenvalue support is usually comprised of one or two cuts of the multi-valued
function $y(z)$ among endpoints. We will see that already in the vacuum configuration of the GWW
model this may or may not be so.

Equations~(\ref{eq:sys1})--(\ref{eq:sys4}) constitute a system of four equations that can be solved
for $a_1$, $a_2$, $b_1$ and $b_2$ as functions of $\lambda$ (a piece of data) and of the parameter $c$
(which needs also to be determined). In fact, the system can be solved explicitly, and we anticipate that
from this point of view there are two kinds of solutions: those in which the support has only one arc, for
which the system itself determines the value of $c$, and those in which the support has two or three arcs,
for which the value of $c$ will be determined by fixing the fraction $q$ of the total unit charge in one of
the arcs (it turns out that in the case of three arcs two of them have the same fraction of the charge).
\section{The GWW model in the vacuum configuration\label{sec:vac}}
Before studying the general structure of the GWW model, in this brief Section we discuss the well-known
results for the vacuum configuration as illustrative  particular solutions of our general approach.
These examples serve to set the pattern of how to calculate the function $y(z)$ and how to calculate
the support $\gamma$ of the eigenvalue density in two different cases. In this section we assume that the integration
path $\Gamma$ is the negatively-oriented unit circle.
\subsection{The weak-coupling regime $0<\lambda<2$}
This case corresponds to solutions of the system (\ref{eq:sys1})--(\ref{eq:sys4}) with $b_1 = b_2 = -1$.
The system can be solved by elementary means and, except for the naming of the roots ($a_1=a$ and $a_2=\bar{a}$
or vice versa), there is a unique solution for $y(z)^2$ given by
\begin{equation}
	\label{ew}
	y(z)^2 = \frac{(z+1)^2(z-a)(z-\bar{a})}{\lambda^2 z^4},
\end{equation}
where
\begin{equation}
	\label{eq:a}
	a = 1-\lambda - i \sqrt{\lambda(2-\lambda)} = e^{-i2\arcsin\left(\sqrt{\lambda/2}\right)},
\end{equation}
and
\begin{equation}
	c=-1+\frac{4}{\lambda}.
\end{equation} 
Note that the restriction to $0<\lambda<2$ guarantees that $a$ is on the unit circle, i.e., that $a\bar{a}=1$.
Note also that the system itself fixes the value of $c$ without additional information.

In this case
\begin{equation}
	\label{ew1}
	y(z) = \frac{(z+1)\sqrt{(z-a)(z-\bar{a})}}{\lambda z^2},
\end{equation}
and the eigenvalue support $\gamma$ is the anti-Stokes line~\cite{AL13}
\begin{equation}
	\re\left(\int_a^z y(z) dz\right) = 0
\end{equation}
joining $\bar{a}$ with $a$ in the negative sense, which turns out to be the corresponding arc of the unit circle.

Equation~(\ref{mes}) gives the explicit formula for the density,
\begin{equation}
	\label{edw}
	\rho^{(\mathrm{w})}(z) = \frac{2}{\pi \lambda }\cos \frac{\theta}{2}\sqrt{\frac{\lambda}{2}-\sin^2\frac{\theta}{2}},
	\quad z=e^{-i\theta},
\end{equation}
and the total electrostatic energy can be calculated using Eq.~(\ref{total}),
\begin{equation}
	\label{tew}
	\mathcal{E}^{(\mathrm{w})} = \frac{3}{4}-\frac{2}{\lambda}-\frac{1}{2}\ln \frac{\lambda}{2}.
\end{equation}
\subsection{The strong-coupling regime $2<\lambda$}
This case corresponds to solutions of the system (\ref{eq:sys1})--(\ref{eq:sys4}) with $a_1=b_1 = a_+$ and
$a_2 = b_2 = a_-$. Again, the system can be solved by elementary means and, except for the naming of the roots,
there is a unique solution for $y(z)^2$ given by
\begin{equation}
	\label{es}
	y(z)^2 = \frac{(z-a_+)^2(z-a_-)^2}{\lambda^2 z^4},
\end{equation}
where
\begin{equation}
	\label{eq:apm}
	a_{\pm} = -\frac{\lambda}{2} \pm \sqrt{\frac{\lambda^2}{4}-1},
\end{equation}
and
\begin{equation}
	c = \frac{4}{\lambda^2}.
\end{equation} 
Note that in this regime both $a_+$ and $a_-$ are real, negative and are placed symmetrically with respect to the unit circle,
i.e., $a_+ a_- = 1$. Note also that Eq.~(\ref{es}) is a ``perfect square'', but the condition given in Eq.~(\ref{rr2}) fixes
the choice of sign and leads to the sectionally analytic function
\begin{equation}\label{es1}
	y(z)
	=
	\left\{\begin{array}{l}
	\displaystyle
	y_+(z) = -\frac{1}{\lambda z^2} (z^2+\lambda z+1), \quad  \mbox{for $|z|>1$},
	\\
	\\
	\displaystyle
	y_-(z) = \frac{1}{\lambda z^2} (z^2+\lambda z+1), \quad \mbox{for $|z|<1$}.
	\end{array}\right.
\end{equation}

Thus we get an eigenvalue density supported on the whole unit circle,
\begin{equation}
	\label{eds}
	\rho^{(\mathrm{s})}(z)
	=
	\frac{1}{2\pi }\left(1+\frac{2}{\lambda}\cos \theta\right) ,\quad z=e^{-i\theta},
\end{equation}
and from Eqs.~(\ref{total}) and~(\ref{eds}) it follows that the total electrostatic energy is
\begin{equation}
	\label{tes}
	\mathcal{E}^{(\mathrm{s})}=-\frac{1}{\lambda^2}.
\end{equation}
Incidentally, Eqs.~(\ref{tew}) and~(\ref{tes}) lead to the well-known third-order
phase transition of the GWW model at the critical value $\lambda=2$.
\section{The general structure of the GWW model\label{sec:gs}}
The expressions of $y(z)$ for both the weak and strong coupling situations have zeros outside the support
of the eigenvalue density ($z=-1$ in the weak coupling case and $z=a_{\pm}$ in the strong coupling case).
Thus the electric field vanishes at these points and we may expect the existence of
configurations in the large-$N$ limit of the GWW model  with cuts developed around these points.
We will show that these configurations correspond to the nonvacuum saddles found numerically for finite $N$
in Ref.~\cite{BD16}. From a mathematical point of view these configurations require a choice of the integration
path $\Gamma$ of the GWW model to include not only the unit circle, but also the negative real axis.
\subsection{General solution of the system of equations}
The system (\ref{eq:sys1})--(\ref{eq:sys4}) admits solutions that are symmetric with respect to the unit circle,
i.e., solutions for which
\begin{equation}
	\label{red}
	b_1=\frac{1}{a_1},\quad b_2 = \frac{1}{a_2}.
\end{equation}
Incidentally, for comparison of our results with the numerical results of Ref.~\cite{BD16}, note that the authors
use in effect an extension of their variable $\theta$ with range $-\pi<\theta<\pi$ to a complex plane $z'$ that
is related to our variable $z$ by
\begin{equation}
	\label{zp}
	z' = -i \log z.
\end{equation}
E.g., symmetry of points in the negative real axis with respect to the unit circle $|z|=1$
translates into symmetry with respect to the origin of the vertical axis in FIG.~1 of Ref.~\cite{BD16}.

Returning to the calculation of solutions with the symmetry given by Eq.~(\ref{red}), if we define
\begin{equation}
	\label{bA}
	A_1 = a_1+\frac{1}{a_1},
	\quad
	A_2 = a_2+\frac{1}{a_2},
\end{equation}
then the system (\ref{eq:sys1})--(\ref{eq:sys4}) reduces to 
\begin{eqnarray}
	A_1+A_2 & = & -2 \lambda, \\ 
	A_1 A_2 & = & (c+1)\lambda^2-4.
\end{eqnarray}
As a consequence,  both $A_1$ and $A_2$ are the solutions of the quadratic equation.
\begin{equation}
	\label{bAA}
	A^2+2 \lambda A+(c+1) \lambda^2 -4=0,
\end{equation}
which implies that $a_1$, $1/a_1$, $a_2$ and $1/a_2$ are the solutions of the quartic equation
\begin{equation}
	\label{bAAA}
	a^4 + 2\lambda (a^3+a) + ((c+1)\lambda^2-2) a^2 +1 = 0.
\end{equation}
These solutions are given by
\begin{eqnarray}
	\label{sole1}
	a_1,\,\frac{1}{a_1}&=&\frac{1}{2}\left(-\lambda+\sqrt{4-c\lambda^2}\pm \sqrt{(1-c)\lambda^2-2\lambda \sqrt{4-c\lambda^2}} \right),\\
	\label{sole2}
	a_2,\,\frac{1}{a_2}&=&\frac{1}{2}\left(-\lambda-\sqrt{4-c\lambda^2}\pm \sqrt{(1-c)\lambda^2+2\lambda \sqrt{4-c\lambda^2}} \right).
\end{eqnarray} 
In fact, it can be checked that, except for the naming of the roots, these are all the 24 solutions of the system
(\ref{eq:sys1})--(\ref{eq:sys4}).

To discuss these roots we eliminate $a$ between Eq.~(\ref{bAAA}) and its derivative with respect to $a$, and find that the critical values
of $c$ at which roots coalesce are given by
\begin{equation}
	\label{crite}
	c = -1 \pm \frac{4}{\lambda},\, \frac{4}{\lambda^2}.
\end{equation}
For example, the non-degenerate configurations in the physical region $\lambda>0$ can be classified as follows:
(i) There are four real solutions if
(a) $0<\lambda<2$ and $c<-1-4/\lambda$,
or
(b) $\lambda>2$ and $c<-1-4/\lambda$ or $-1+4/\lambda<c<4/\lambda^2$;
(ii) There are four complex solutions if
(a) $\lambda>0$ and $c>4/\lambda^2$
or
(b) $0<\lambda<2$ and $-1+4/\lambda<c<4/\lambda^2$;
(iii) The remaining cases correspond to two real solutions and two complex solutions.
\subsection{Eigenvalue support configurations}
\begin{figure}
	\includegraphics{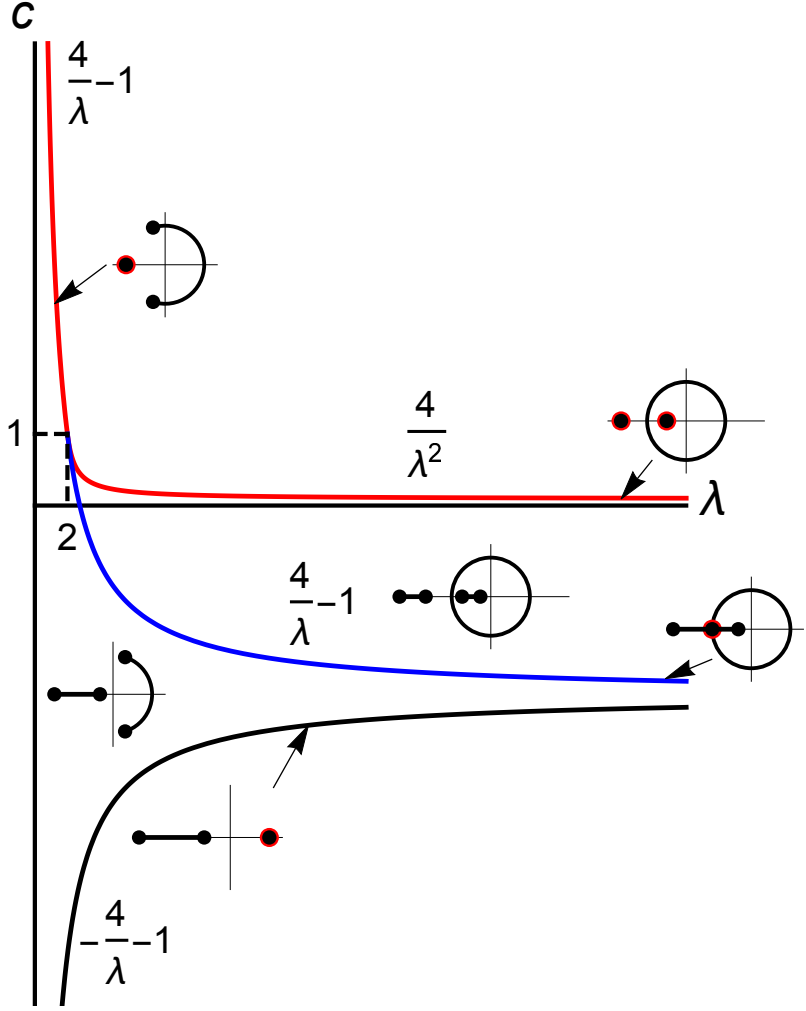}
	\caption{\label{fig:c} (Color online) Schematic representation of the acceptable configurations for the eigenvalue
	                                 support of the GWW model in the right half of the $(\lambda,c)$ plane.}
\end{figure}

To find the possible acceptable eigenvalue densities we have to study the differential
\begin{equation}
	\label{gena}
	y(z) dz = \frac{\sqrt{(z-a_1)(z-a_2)(z-1/a_1)(z-1/a_2)}}{\lambda z^2}\,dz
\end{equation}
in each of the regions (i)--(iii) and on their borders (where double roots may cancel the square root as we saw, for instance,
in the strong coupling regime of the vacuum configuration) to see if Eq.~(\ref{mes}) leads to a unit-normalized eigenvalue density
and to calculate its support. A detailed discussion of the techniques we have used to carry out this calculations
can be found in Refs.~\cite{AL13,AL15,AL15b}, and we have summarized the results of this study schematically
in FIG.~\ref{fig:c}.

The right half of the $(\lambda,c)$ plane is divided into four regions by the curves $c=-4/\lambda-1$
(the lowest, black curve in FIG.~\ref{fig:c}), $c=4/\lambda-1$ (the middle, blue curve in FIG.~\ref{fig:c}),
and the piecewise-defined curve $c=4/\lambda-1$ for $0<\lambda\leq 2$ and $c=4/\lambda^2$ for $\lambda>2$
(the upper, red curve in FIG.~\ref{fig:c}).

In the regions below the first curve and above the third curve there are not any acceptable configurations.
The region between the first and second curves is a two-cut region, where the eigenvalue support consists of
an arc of the unit circle and a segment of the negative real axis. And in the region between the second and third
curves the support consists of three arcs: the whole unit circle and two intervals on the negative real axis symmetrically
placed with respect to the unit circle. The configurations corresponding to the critical lines are marked with arrows
pointing to the corresponding parts of the line. For example, along the first, lowest curve, the eigenvalue support
consists of a single cut on the negative real axis (there is also a double zero at $z=1$).
And the vacuum configurations discussed in the previous Section correspond to the third curve (marked
in red in FIG.~\ref{fig:c}): for $0<\lambda<2$  (the weak coupling region) the support is an arc of the unit circle
and for $\lambda>2$ (the strong coupling region) the support is the whole unit circle. Note also that the critical
value $\lambda=2$ corresponds to the four roots $a_1=a_2=b_1=b_2=-1$. Finally, the critical blue line in
FIG.~\ref{fig:c} corresponds to configurations with a support consisting of the whole unit circle plus an intersecting
interval of the negative real axis (in these configurations $-1$ is a double root).

The region on nonvacuum saddles at weak coupling of Ref.~\cite{BD16} is the region above the
black curve, below the red curve and to the left of $\lambda=2$. In this case, for points $z=x$ on the real
cut $[a_1,1/a_1]$, Eq.~(\ref{gena}) reduces to
\begin{equation}
	y(x)dx = \frac{dx}{\lambda x} |x-a_2| \sqrt{(x-a_1)(x-1/a_1)}.
\end{equation}
Note that the argument inside the square root is negative, and therefore Eq.~(\ref{mes}) leads to
a positive definite eigenvalue density. For points $z=e^{-i\theta}$ on the unit circle,
Eq.~(\ref{gena}) takes the form
\begin{equation}
	y(z) dz = \frac{2\,d\theta}{i\lambda}
	               \sqrt{\left(\cos\theta-\frac{A_1}{2}\right)\left(\cos\theta-\frac{A_2}{2}\right)},
\end{equation}
and the condition of positivity of $\rho(z)$ determines the arc of the unit circle that supports the density.

Likewise, the region of nonvacuum saddles at strong coupling of Ref.~\cite{BD16} is the region above
the blue curve and below the red curve (and therefore to the right of $\lambda=2$). In this case we can use
similar arguments with the slight variations that there are two real intervals and the complete unit circle.

The unit normalization of all these acceptable eigenvalue densities is a consequence of Eq.~(\ref{eq:sys1}),
and the relation between $c$ and the eigenvalue density given in Eq.~(\ref{cs1}) is a consequence of
Eq.~(\ref{eq:sys2}). Incidentally, for those instances in which the whole unit circle is part of the
eigenvalue support, opposite signs of the square root in Eq.~(\ref{gena}) have to be taken inside and
outside the unit circle. (A similar prescription has been used in Eq.~(\ref{es1}).)
\subsection{Eigenvalue support configurations as a function of the charge fraction}
Although the description of the support configurations schematized in FIG.~\ref{fig:c} is complete, it is given
in terms of the parameter $c$, which does not permit a direct comparison with the results of Ref.~\cite{BD16},
that are given in terms of the transferring of eigenvalues among the different arcs of the support.

\begin{figure}
	\includegraphics{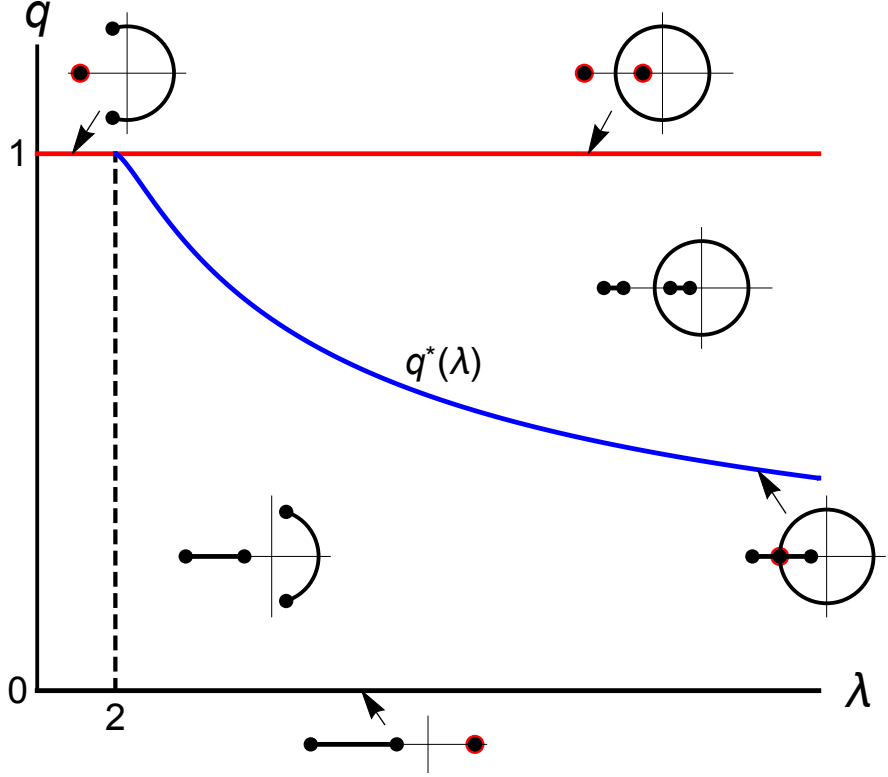}
	\caption{\label{fig:q} (Color online) Schematic representation of the acceptable configurations for the eigenvalue
	                                  support of the GWW model in the $(\lambda,q)$ plane, where $q$ is the fraction
	                                  of the total unit charge that is supported on (part or the whole) unit circle.}
\end{figure}

To remedy this shortcoming, in FIG.~\ref{fig:q} we schematize the eigenvalue support configuration
as a function of $\lambda$ and of the fraction $q$ of the total unit charge that is supported on (part or the whole)
unit circle. This prescription might seem ambiguous in the three-arc case, where the support
consists of the two real cuts $[a_1,a_2]$ and $[1/a_2,1/a_1]$ and the whole unit circle, but
the charge fractions on the two real cuts are equal. Indeed, using the change of variable $x\rightarrow 1/x$
and recalling the definitions of $A_1$ and $A_2$ given in Eq.~(\ref{bA}) we get,
\begin{eqnarray}
	\int_{a_1}^{a_2}\rho(x)\,dx
	& = &
	\frac{1}{2\pi \lambda}\int_{a_1}^{a_2}\sqrt{\left|(x^2-A_1x+1)(x^2-A_2 x+1)\right|}\,\frac{dx}{ x^2}\\
	& = &
	\frac{1}{2\pi \lambda}\int_{1/a_2}^{1/a_1}\sqrt{\left|(x^2-A_1x+1)(x^2-A_2 x+1)\right|}\,\frac{ d  x}{x^2}\\
	& = &
	\int_{1/a_2}^{1/a_1}\rho(x)\,dx,
\end{eqnarray}
and the fraction of the charge supported on the whole unit circle is given by
\begin{equation}
	\label{uc}
	q = \frac{2}{\pi \lambda}\int_0^{\pi }\sqrt{\cos^2 \theta+\lambda \cos\theta+(c+1)(\lambda/2)^2-1} \, d \theta.
\end{equation}
Note also that
\begin{equation}
	\label{dq}
	\frac{\partial q}{\partial c}=\frac{1}{4\pi^2}\int_0^{\pi}\frac{ d  \theta}{\rho(\theta)}>0.
\end{equation}
If we move from the curve $c=4/\lambda^2$ to the critical point $c=-1+4/\lambda$ at constant $\lambda$,
the endpoints $1/a_1$ and $a_1$ of the two real cuts coalesce and the gap between these two cuts closes.
Moreover, from Eq.~(\ref{uc}) we obtain that the critical charge on the circle when the gap closes is
\begin{eqnarray}
	q^*(\lambda)
	& = &\frac{2}{\pi \lambda}\int_0^{\pi }\sqrt{\cos^2 \theta+\lambda \cos\theta+\lambda-1} \, d\theta\\
	& = & \frac{2}{\pi \lambda}
	         \left[\sqrt{2\lambda-4}+\lambda\arctan\left(\sqrt{\frac{2}{\lambda-2}}\right) \right].
	\label{cc}
\end{eqnarray}
As an illustration of this result we apply Eq.~(\ref{cc}) to the numerical example of Ref.~\cite{BD16}:
with $N=40$ and $\lambda=4$, we might expect about $(1-q^*(4)) 40 \approx 7.3$ eigenvalues
on the vertical axis, in good agreement with the value $m=7$ of FIG.~1(i) in Ref.~\cite{BD16}.

The colors of the curves in FIG.~\ref{fig:q} match those of the curves in FIG.~\ref{fig:c}. Thus,
the vacuum configuration corresponds to $q=1$ in FIG.~\ref{fig:q}, and the lowest curve
$c=-1-4/\lambda$ in FIG.~\ref{fig:c} maps to the horizontal line $q=0$ in FIG.~\ref{fig:q}.
\subsection{Instanton actions in the weak- and strong-coupling regions}
In this Section we will address the interpretation of the migration of eigenvalues from the unit
circle to one or the two cuts on the negative real axis as eigenvalue tunneling. The weak-coupling case
has been calculated directly, but the strong-coupling action required a trans-series ansatz in the
string equation~\cite{MA08b}. We will see that by implementing the physical interpretations put forward
in Ref.~\cite{BD16} in our formalism, we recover both results on an equal footing.

The two-cut eigenvalue configurations can be interpreted in terms of a tunneling process in which
eigenvalues from the one-cut weak-coupling phase case lying on the unit circle with endpoints $a$
and $\bar{a}$ given in Eq.~(\ref{eq:a}) migrate to the cut containing the electrostatic equilibrium point $z=-1$.
Thus, for the tunneling of a small filling fraction $\delta q$ we have
\begin{equation}
	\label{w1}
	\frac{ Z_n[ \rho^{(\mathrm{w})} + \delta \rho^{(\mathrm{w})} ]}{ Z_n[ \rho^{(\mathrm{w})} ]}
	=
	\exp{\left(-N^2 \int_{\gamma}  d  z \frac{\delta S}{\delta\rho(z)}[\rho^{(\mathrm{w})}]\, \delta\rho^{(\mathrm{w})}
	+ o(N^2)\right)}, \quad \mbox{as $N\rightarrow \infty$},
\end{equation}
where
\begin{equation}
	\delta\rho^{(\mathrm{w})}
	=
	 \left(\delta(z+1)-\delta(z-a)\right)\delta q.
\end{equation}
Hence, recalling Eq.~(\ref{hpo}) and the definition of the total potential $U(z)=\re(W_\mathrm{h}(z))$, we obtain that
\begin{equation}
	\re\left(S[\rho^{(\mathrm{w})}+\delta\rho^{(\mathrm{w})}]-S[\rho^{(\mathrm{w})}]\right)
	\approx \left(U(-1)-U(a) \right)\delta q
	= S_\mathrm{I}^{(\mathrm{w})}\delta q,
\end{equation}
where $S_\mathrm{I}^{(\mathrm{w})}$ is the weak-coupling instanton action. Therefore,
\begin{eqnarray}
	S_\mathrm{I}^{(\mathrm{w})}
	& = &
	U(-1)-U(a) \\
	& = &
	\re \int_{a}^{-1} y(z)  dz\\
	& = & 
	\frac{1}{\lambda}\,\int_{a}^{-1} (z+1) \sqrt{(z-a)(z-a^*)}\,\frac{dz}{z^2},
\end{eqnarray}
and a straightforward calculation yields (see Eq.~(4.27) in Ref.~\cite{MA08b})
\begin{equation}
	\label{winsta}
	S_\mathrm{I}^{(\mathrm{w})}
	=
	\frac{4}{\lambda}\sqrt{1-\frac{\lambda}{2}}-\arccosh\left(\frac{4}{\lambda}-1 \right).
\end{equation}

Likewise, the three-cut eigenvalue distribution can be interpreted in terms of a tunneling process
in which pairs of eigenvalues from the one-cut strong case lying on the unit circle tunnel to the cuts
around the equilibrium points $a_{\pm}$ given in Eq.~(\ref{eq:apm}). Thus, in this case we have an expression
similar to Eq.~(\ref{w1}) with $\rho^{(\mathrm{w})}$ replaced by $\rho^{(\mathrm{s})}$ and with
\begin{equation}
	\label{drh}
	\delta\rho^{(\mathrm{s})}(z)
	=
	\left(\delta(z-a_+)+\delta(z-a_-)-2 \delta(z+1)\right)\delta q.
\end{equation}
Hence we obtain
\begin{eqnarray}
	\re \left(S[\rho^{(\mathrm{s})}+\delta\rho^{(\mathrm{s})}]-S[\rho^{(\mathrm{s})}]\right)
	& = &
	 \left(U(a_+)+U(a_-)-2U(-1) \right)\delta q,\\
	& = &
	\re\left( \int_{-1}^{a_+} y(z_+)  d  z+ \int_{-1}^{a_-} y(z_-)  d z\right)\delta q.
\end{eqnarray}
Using again Eqs.~(\ref{es1}) and~(\ref{eq:apm}), taking into account that $a_+a_-=1$, we easily obtain
\begin{equation}
	\re \int_{-1}^{a_{\pm}} y(z_{\pm})  dz
	=
	\frac{1}{\lambda}\left( a_+-\frac{1}{a_+}\right)+ \log |a_+|
 	=
	-\sqrt{1-\frac{4}{\lambda^2}}+\arccosh\left(\frac{\lambda}{2}\right).
 \end{equation}
 Then we have that
 \begin{equation}
	\Big|\re\left(S[\rho^{(\mathrm{s})}+\delta\rho^{(\mathrm{s})}]-S[\rho^{(\mathrm{s})}]\right)\Big|
	\approx  S_\mathrm{I}^{(\mathrm{s})}\delta q,
\end{equation}
where $S_\mathrm{I}^{(\mathrm{s})}$ is
the strong-coupling instanton action 
 \begin{equation}
 	\label {scia}
 	S_\mathrm{I}^{(\mathrm{s})}
	=
	2\,\arccosh\left(\frac{\lambda}{2}\right)-2\sqrt{1-\frac{4}{\lambda^2}},
 \end{equation}
 in agreement with the analytic expression given in Eq.~(4.36) of
 Ref.~\cite{MA08b}, and with the numerical result showed by the blue line in Fig.~4 of Ref.~\cite{BD16}.
\subsection{The $q=0$ configuration}
We finally mention that the $q=0$ configuration in FIG.~\ref{fig:q}, where the whole eigenvalue support
is an interval on the negative real axis, can be treated in complete analogy to the vacuum configurations
discussed in Sec.~\ref{sec:vac}. Note that to reach this limiting configuration it is essential to use the choice
of $\Gamma$ mentioned in the first paragraph of Sec.~\ref{sec:gs}.

This case corresponds to solutions of the system (\ref{eq:sys1})--(\ref{eq:sys4})
with $a_1 = a_2 = 1$. The system can be solved by elementary means and, except for the naming of the roots,
there is a unique solution for $y(z)^2$ given by
\begin{equation}
	y(z)^2 = \frac{(z-1)^2(z-b_+)(z-b_-)}{\lambda^2 z^4},
\end{equation}
where
\begin{equation}
	b_\pm = -1-\lambda \pm \sqrt{\lambda(2+\lambda)},
\end{equation}
and, as FIG.~\ref{fig:c} illustrates,
\begin{equation}
	c=-1-\frac{4}{\lambda}.
\end{equation} 

The eigenvalue support $\gamma$ is the $[b_-,b_+]$ interval of the negative real axis,
Eq.~(\ref{mes}) gives the explicit formula for the density,
\begin{equation}
	\rho^{(0)}(z) = \frac{(1-x)\sqrt{(b_+-x)(x-b_-)}}{2\pi \lambda x^2},
	\quad x\in[b_-,b_+],
\end{equation}
and the total electrostatic energy can be calculated using Eq.~(\ref{total}),
\begin{equation}
	\label{te0}
	\mathcal{E}^{(0)} = \frac{3}{4}+\frac{2}{\lambda}-\frac{1}{2}\ln\frac{\lambda}{2}.
\end{equation}
\section{Summary}
The rich variety of configurations of saddle points in the Gross-Witten-Wadia matrix model revealed by the
detailed numerical calculations of Buividovich, Dunne, and Valgushev in Ref.~\cite{BD16} can be completely
justified in the large-$N$ limit by using techniques similar to those developed for the determination of the
asymptotic support of the zeros of certain non-hermitian families of orthogonal polynomials~\cite{AL13,AL15b,AL15}.

These methods are a generalization of those used for hermitian families of orthogonal polynomials,
wherein one attempts to write a system of equations for the candidate endpoints of the support,
solves the system, and calculates the density that is supported in one or more intervals that are
cuts of a multi-valued function $y(z)$. The success of the generalization of this strategy to the GWW
relies on using directly the matrix eigenvalues as integration variables, which in turn renders the resulting
system explicitly solvable. We also remark that by implementing in our formalism the physical interpretation
of nonvacuum saddles in terms of eigenvalue tunneling put forward in Ref.~\cite{BD16} we have been able
to calculate by the same method both the weak-coupling and strong-coupling instanton actions that were
formerly obtained analytically in Ref.~\cite{MA08b} using trans-series solutions to the recursion relations
characterizing the free energy.
\begin{acknowledgments}
We wish to acknowledge the support of the Spanish Ministerio de Econom\'{\i}a y Competitividad under
Project No.~FIS2015-63966-P.
\end{acknowledgments}
%
\end{document}